# A semi-relativistic treatment of spinless particles subject to the nuclear Woods-Saxon potential

M. Hamzavi[1*], S. M. Ikhdair[2**], A. A. Rajabi[3]

[1]Department of Basic Sciences, Shahrood Branch, Islamic Azad University, Shahrood, Iran

[2]Physics Department, Near East University, 922022 Nicosia, North Cyprus, Mersin 10, Turkey

[3]Physics Department, Shahrood University of Technology, Shahrood, Iran

[*] Corresponding author: Tel.:+98 273 3395270, fax: +98 273 3395270
[*] majid.hamzavi@gmail.com
[**] sikhdair@neu.edu.tr

**Abstract:**

By applying an appropriate Pekeris approximation to deal with the centrifugal term, we present an approximate systematic solution of the two-body spinless Salpeter (SS) equation with the Woods-Saxon interaction potential for arbitrary $l$-state. The analytical semi-relativistic bound-state energy eigenvalues and the corresponding wave functions are calculated. Two special cases from our solution are studied: the approximated Schrödinger-Woods-Saxon problem for arbitrary $l$-state and the exact $s$-wave ($l=0$).

## 1. Introduction

The Bethe–Salpeter (BS) equation [1], named after Hans Bethe and Edwin Salpeter, describes the bound states of a two-body (particles) quantum field theoretical system in a relativistic covariant formalism. The equation was actually first published in 1950 at the end of a paper by Yoichiro Nambu but without derivation [2].

Due to its generality and its application in various branches of theoretical physics, the BS equation appears in many different forms [3-10]. Different aspects of this equation have been elegantly studied mainly by Lucha *et al.* in [11-25] and some other authors where they have worked on many interesting approaches to deal with its nonlocal Hamiltonian [26-30].

The Woods-Saxon (WS) potential is a short range potential and widely used in nuclear, particle, atomic, condensed matter and chemical physics [31-39]. This potential is reasonable for nuclear shell models and used to represent the distribution of nuclear densities. The WS and spin-orbit interaction are important and applicable to deformed nuclei [40] and to strongly deformed nuclides [41]. The WS potential parameterization at large deformations for plutonium $^{237,239,241}Pu$ odd isotopes was analyzed [33]. The structure of single-particle states in the second minima of $^{237,239,241}Pu$ has been calculated with an exactly WS potential. The Nuclear shape was parameterized. The parameterization of the spin-orbit part of the potential was obtained in the region corresponding to large deformations (second minima) depending only on the nuclear surface area. The spin-orbit interaction of a particle in a non-central self-consistent field of the WS type potential was investigated for light nuclei and the scheme of single-particle states has been found for mass number $A_0 = 10$ and 25 [40]. Two parameters of the spin-orbit part of the WS potential, namely the strength parameter and radius parameter were adjusted to reproduce the spins for the values of the nuclear deformation parameters [42].

The usual WS potential takes the form [31]

$$V(r) = -\frac{V_0}{1+e^{\frac{r-R}{a}}}, \quad (1)$$

where $V_0$ is the depth of potential, $a$ is the diffuseness of the nuclear surface and $R$ is the width of the potential [31, 38-39].

The aim of the present work is to study the usual WS potential within the framework of a semi-relativistic SS equation and obtain an approximate bound-state energy eigenvalues and their corresponding wave functions. We use a simple and powerful tool in the form of a parametric generalization of the Nikiforov-Uvarov (NU) method [43]. Such a shortcut of the method has proved its effectiveness in solving various potential models over the past few years [44].

The present work is organized as follows. In Section 2, we review the SS equation and apply it to the usual WS potential interaction to obtain the semi-relativistic SS bound-state energy spectrum and their corresponding wave functions for two-interacting particles. In Section 3, we consider the solution for the non-relativistic case. Finally, in Section 4 we give our final comments and conclusion.

## 2. Spinless Salpeter equation and its application to Woods-Saxon potential

The SS equation for a two-body system under a spherically symmetric potential in the center-of-mass system has the form [26, 27, 29]

$$\left[\sum_{i=1,2}\sqrt{-\Delta+m_i^2}+V(r)-M\right]\chi(\vec{r})=0, \quad \Delta=\nabla^2, \quad (2)$$

where the kinetic energy terms involving the operation $\sqrt{-\Delta+m_i^2}$ are non-local operators and $\chi(\vec{r})=\psi_{nl}(r)Y_{lm}(\theta,\phi)$ stands for the semi-relativistic wave function. For heavy interacting particles, the kinetic energy operators in Eq. (2) can be approximated as [2-32]

$$\sqrt{-\Delta+m_i^2}=m_1+m_2-\frac{\Delta}{2\mu}-\frac{\Delta^2}{8\eta^3}-..., \quad (3)$$

where $\mu=m_1m_2/(m_1+m_2)$ stands for the reduced mass and $\eta=\mu\left(m_1m_2/(m_1m_2-3\mu^2)\right)^{1/3}$ is simply an introduced useful mass parameter [25,26]. The above Hamiltonian containing relativistic corrections up to order $(v^2/c^2)$ and is called a generalized Breit-Fermi Hamiltonian [11-15]. Using an appropriate transformation and following the same procedures explained in Ref. [29] (see Eqs. (13)- (18)), one can then arrive at the semirelativistic SS equation:

$$\left[-\frac{\hbar^2}{2\mu}\frac{d^2}{dr^2}+\frac{l(l+1)\hbar^2}{2\mu r^2}+W_{nl}(r)-\frac{W_{nl}^2(r)}{2\tilde{m}}\right]\psi_{nl}(r)=0, \quad (4)$$

where

$$W_{nl}(r)=V(r)-E_{nl}, \text{ and } \tilde{m}=\frac{\eta^3}{\mu^2}=\frac{m_1m_2\mu}{m_1m_2-3\mu^2}. \quad (5)$$

Now, we intend to solve the above semi-relativistic equation (4) with the usual Woods-Saxon potential interaction (1). Thus, the insertion of Eq. (1) into (5) allows us to obtain

$$\left\{\frac{d^2}{dr^2}-\frac{l(l+1)}{r^2}+\frac{2\mu}{\hbar^2}\left[E_{nl}\left(1+\frac{E_{nl}}{2\tilde{m}}\right)+V_0\left(1+\frac{E_{nl}}{\tilde{m}}\right)y+\frac{V_0^2}{2\tilde{m}}y^2\right]\right\}\psi_{nl}(r)=0,$$

$$y=\frac{1}{1+e^{vx}}, \quad v=\frac{1}{a}, \quad x=r-R. \quad (6)$$

Because Eq. (6) cannot be solved analytically due to the centrifugal term $l(l+1)r^{-2}$, we have to use a proper approximation of this term. Unlike the usual approximation used for the first time in Greene and Aldrich work [45], here we apply the Pekeris approximation by taking an expansion around $r = R$ (or $x = 0$) in series of powers of $x/R$ as [46]:

$$V_l(r) = \frac{l(l+1)}{r^2} = \frac{l(l+1)}{R^2\left(1+\frac{x}{R}\right)^2} \cong \frac{l(l+1)}{R^2}\left[1 - 2\left(\frac{x}{R}\right) + 3\left(\frac{x}{R}\right)^2 + \ldots\right]. \quad (7)$$

Here the first three terms should be sufficient. Further, the centrifugal term can also be replaced by the usual Woods-Saxon potential form:

$$\tilde{V}_l(r) = \frac{l(l+1)}{r^2} \cong \frac{l(l+1)}{R^2}\left(D_0 + D_1 y + D_2 y^2\right), \quad (8)$$

where $D_i$ ($i = 0,1,2$) can be determined as a function of specific potential parameters [34]. If we expand the expression (8) around $r = R$ (or $x = 0$) up to the second-order term $(x/R)^2$ and next compare it with Eq. (7), we can finally obtain the explicit forms of the parameters $D_0$, $D_1$ and $D_2$ as

$$D_0 = 1 - \frac{4}{\nu R} + \frac{12}{\nu^2 R^2}, \quad (9a)$$

$$D_1 = \frac{8}{\nu R} - \frac{48}{\nu^2 R^2}, \quad (9b)$$

$$D_2 = \frac{48}{\nu^2 R^2}. \quad (9c)$$

Thus, we can replace the centrifugal term (7) by its approximation (8) to obtain an approximate analytical solution for Eq. (6) as

$$\left\{\frac{d^2}{dx^2} + \frac{1}{(1+e^{\nu x})}\left[\frac{2\mu V_0}{\hbar^2}\left(1 + \frac{E_{nl}}{\tilde{m}}\right) - \frac{l(l+1)D_1}{R^2}\right]\right.$$

$$\left. + \frac{1}{(1+e^{\nu x})^2}\left(\frac{2\mu V_0^2}{2\hbar^2 \tilde{m}} - \frac{l(l+1)D_2}{R^2}\right) + \left[\frac{2\mu E_{nl}}{\hbar^2}\left(1 + \frac{E_{nl}}{2\tilde{m}}\right) - \frac{l(l+1)D_0}{R^2}\right]\right\}\psi_{nl}(r) = 0. \quad (10)$$

Hence, the above equation is amendable to the solution of the NU method [43]. Let us now introduce an appropriate transformation $s = \frac{1}{1+e^{\nu x}}$ into Eq. (10) to recast it in a more simple form:

$$\left[\frac{d^2}{ds^2}+\frac{1-2s}{s(1-s)}\frac{d}{ds}+\frac{1}{v^2 s^2(1-s)^2}\left(-As^2+Bs-C\right)\right]\psi_{nl}(s)=0, \quad (11)$$

where

$$A=\frac{1}{v^2}\left(\frac{l(l+1)D_2}{R^2}-\frac{2\mu V_0^2}{2\hbar^2\tilde{m}}\right), \quad (12a)$$

$$B=\frac{1}{v^2}\left[\frac{2\mu V_0}{\hbar^2}\left(1+\frac{E_{nl}}{\tilde{m}}\right)-\frac{l(l+1)D_1}{R^2}\right], \quad (12b)$$

$$C=\frac{1}{v^2}\left[\frac{l(l+1)D_0}{R^2}-\frac{2\mu E_{nl}}{\hbar^2}\left(1+\frac{E_{nl}}{2\tilde{m}}\right)\right]. \quad (12c)$$

By comparing Eq. (11) with the Relation (A1), we find the coefficients,

$$c_1=1, \quad c_2=2, \quad c_3=1, \quad (13)$$

and further the relation (A5) determines the rest of the coefficients as

$$c_4=0, \quad c_5=0,$$

$$c_6=A, \quad c_7=-B,$$

$$c_8=C, \quad c_9=A-B+C,$$

$$c_{10}=2\sqrt{C}, \quad c_{11}=2\sqrt{A-B+C},$$

$$c_{12}=\sqrt{C}, \quad c_{13}=\sqrt{A-B+C}, \quad (14)$$

where $A+C>B$ is the essential requirement for the bound state solutions. Therefore, from Relation (A10), we can obtain the binding energy eigenvalue equation as

$$n+\frac{1}{2}+\sqrt{C}+\sqrt{A-B+C}=\frac{1}{2}\sqrt{1+4A}, \quad (15)$$

where

$$A-B+C=\frac{l(l+1)}{v^2 R^2}\left(1+\frac{4}{vR}+\frac{12}{v^2 R^2}\right)-\frac{2\mu}{v^2\hbar^2}\left[E_{nl}+V_0+\frac{(E_{nl}+V_0)^2}{2\tilde{m}}\right], \quad (16a)$$

$$C=\frac{1}{v^2}\left[\frac{l(l+1)}{R^2}\left(1-\frac{4}{vR}+\frac{12}{v^2 R^2}\right)-\frac{2\mu E_{nl}}{\hbar^2}\left(1+\frac{E_{nl}}{2\tilde{m}}\right)\right], \quad (16b)$$

$$1+4A=1+\frac{4}{v^2}\left(\frac{48l(l+1)}{v^2 R^4}-\frac{\mu V_0^2}{\hbar^2\tilde{m}}\right), \quad (16c)$$

for which the binding energy is a negative quantity (i.e., $E_{nl}<0$) and of small value satisfying the inequality $E_{nl}\ll\tilde{m}$ [29]. Obviously, Eq. (15) seems to be a complicated transcendental energy eigenvalue equation admitting two solutions. However, we choose the negative one as mentioned before.

Now, we seek to find the bound-state energy eigenvalues numerically via Eqs. (15) and (16) by taking a set of physical parameter values for $^{208}Pb$ as $m_1 = m_2 = 4.76504\,\text{fm}^{-1}\,(938\,MeV)$, $V_0 = 0.3431032\,\text{fm}^{-1}\,(67.54\,MeV)$, $R = 7.6136\,fm$ and $a = 0.65\,\text{fm}^{-1}$ [38, 39]. Hence, our numerical results are given in Table 1 for various values of radial and orbital quantum numbers $n$ and $l$.

**Table 1** Approximate energy eigenvalues of the SS particles subject to the usual WS potential for various values of $n$ and $l$ quantum numbers.

| $n$ | $l$ | $E_{n,l}$ |
|---|---|---|
| 1 | 0 | -0.345316379 |
| 2 | 0 | -1.007879100 |
|   | 1 | -1.006973500 |
| 3 | 0 | -2.047025224 |
|   | 1 | -2.046833187 |
|   | 2 | -2.046625724 |
| 4 | 0 | -3.656078626 |
|   | 1 | -3.656902452 |
|   | 2 | -3.657700993 |
|   | 3 | -3.658474632 |

To show the behavior of the energy eigenvalues on the usual WS parameters, we involve in plotting the binding energy eigenvalues of SS equation for the usual WS potential versus diffuseness of the nuclear surface $a$ and the width of the potential $R$ in Figures 1 and 2, respectively. As seen from Figure 1, when the diffuseness of the nuclear surface $a$ increases, the energy increases and the binding energy decreases with the increasing of the width of the potential $R$ as shown in Figure 2.

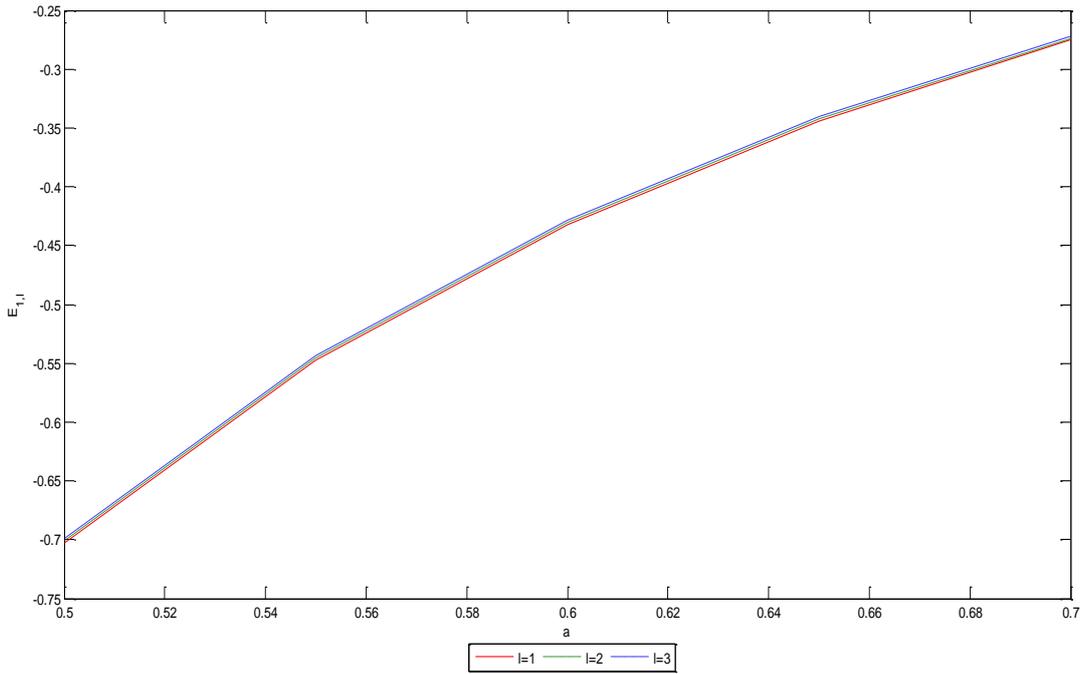

**Fig. 1:** Energy behavior of the SS equation with the usual WS potential versus $a$ for various values of $l$.

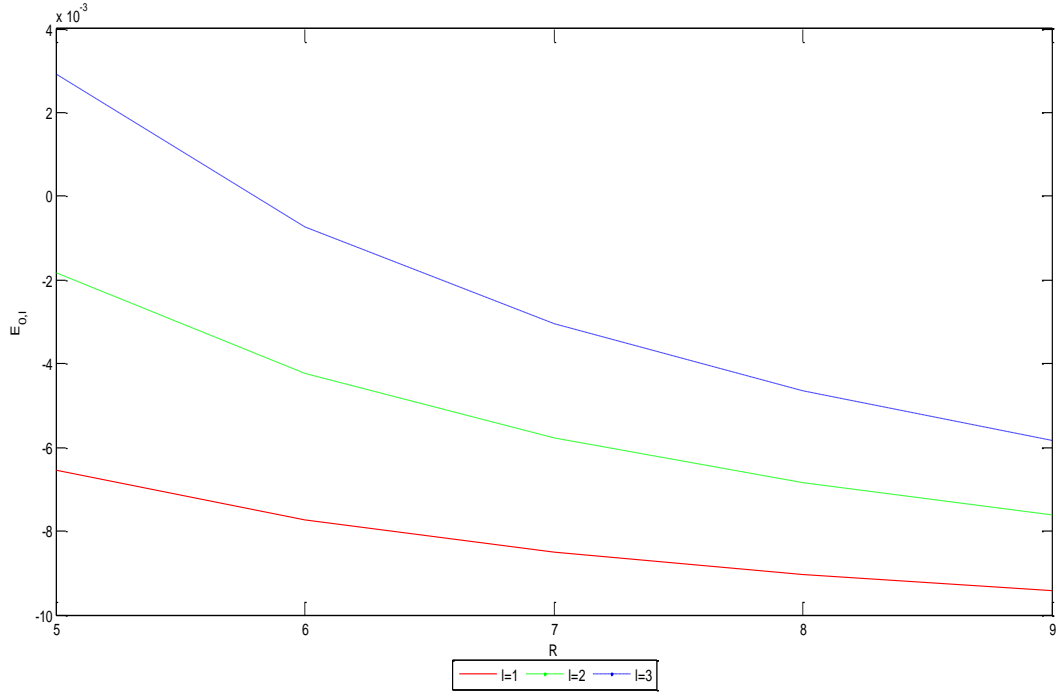

**Fig. 2:** Energy behavior of the SS equation with the usual WS potential versus $R$ for various $n$.

Let us now turn to the calculations of the corresponding wave functions. Referring to Eq. (14) and Relations (A11) and (A12) of Appendix A, we find the functions

$$\rho(s) = s^{2\sqrt{C}}(1-s)^{2\sqrt{A-B+C}}, \quad (17a)$$

$$\phi(s) = s^{\sqrt{C}}(1-s)^{\sqrt{A-B+C}}. \quad (17b)$$

Hence, the relation (A13) gives the first part of the wave functions

$$y_n(s) = P_n^{(2\sqrt{C},2\sqrt{A-B+C})}(1-2s). \quad (18)$$

By using $R_{nl}(s) = \phi(s)y_n(s)$, we get the radial SS wave functions for the usual WS potential from the relation (A14) as

$$\psi_{nl}(s) = A_{nl} s^{\sqrt{C}}(1-s)^{\sqrt{A-B+C}} P_n^{(2\sqrt{C},2\sqrt{A-B+C})}(1-2s), \quad (19)$$

and after substituting $s = \dfrac{1}{1+e^{v(r-R)}}$, we obtain

$$\psi_{nl}(r) = A_{nl}\left(\frac{1}{1+e^{v(r-R)}}\right)^{\sqrt{C}}\left(\frac{e^{v(r-R)}}{1+e^{v(r-R)}}\right)^{\sqrt{A-B+C}} P_n^{(2\sqrt{C},2\sqrt{A-B+C})}\left(\frac{e^{v(r-R)}-1}{e^{v(r-R)}+1}\right), \quad (20)$$

where $A_{nl}$ is the normalization constant. We have $2\sqrt{C} > -1$ and $2\sqrt{A-B+C} > -1$.

## 3. Solution of the non-relativistic case

We consider the Schrödinger for two-body system interacting via the usual WS potential field. Under some limiting conditions when $W_{nl}(r) \ll \tilde{m}$, the semi-relativistic SS equation (4) reduces to the Schrödinger wave equation:

$$\left[\frac{d^2}{dr^2} - \frac{l(l+1)}{r^2} + \frac{2\mu}{\hbar^2}(E_{nl} - V(r))\right]\psi_{nl}(r) = 0. \quad (21)$$

The binding energy equation of Eq. (21) can be easily obtained via Eqs. (15) and (16) as

$$n - L + \sqrt{\frac{l(l+1)}{v^2 R^2}\left(1 - \frac{4}{vR} + \frac{12}{v^2 R^2}\right) - \frac{2\mu E_{nl}}{v^2 \hbar^2}}$$
$$= -\sqrt{\frac{l(l+1)}{v^2 R^2}\left(1 + \frac{4}{vR} + \frac{12}{v^2 R^2}\right) - \frac{2\mu}{v^2 \hbar^2}(E_{nl} + V_0)},$$
$$L(L+1) = \frac{48 l(l+1)}{v^4 R^4}, \qquad L = -\frac{1}{2} + \frac{1}{2}\sqrt{1 + \frac{192 l(l+1)}{v^4 R^4}}, \quad (22)$$

and followed by a lengthy but straightforward algebra, we finally obtain the energy formula for the non-relativistic case as

$$E_{nl} = \frac{\hbar^2 l(l+1)}{2\mu R^2}\left(1 - \frac{4}{vR} + \frac{12}{v^2 R^2}\right) - \frac{\hbar^2}{2\mu(n-L)^2}\left(\frac{4l(l+1)}{v^2 R^3} - \frac{\mu V_0}{v\hbar^2} - \frac{1}{2}(n-L)^2\right)^2. \quad (23)$$

Now, we calculate numerically the non-relativistic energy eigenvalues from Eq. (23) with the aid of Eq. (22) using a set of physical parameter values for $^{208}Pb$ as $m_1 = m_2 = 4.76504 \, \text{fm}^{-1} \, (938 MeV)$, $V_0 = 0.3431032 \, \text{fm}^{-1} \, (67.54 MeV)$, $R = 7.6136 \, fm$ and $a = 0.65 \, fm^{-1}$ [38, 39]. These numerical energies are displayed in Table 2 for various $n$ and $l$ states.

**Table 2** Approximate energy eigenvalues of the Schrodinger particles subject to the usual WS potential for various values of $n$ and $l$ quantum numbers.

| $n$ | $l$ | $E_{n,l}$ |
|---|---|---|
| 1 | 0 | -0.223223240 |
| 2 | 0 | -0.336182678 |
|   | 1 | -0.327759906 |
| 3 | 0 | -0.590280900 |
|   | 1 | -0.581509600 |
|   | 2 | -0.564026223 |
| 4 | 0 | -0.954658926 |
|   | 1 | -0.945425071 |
|   | 2 | -0.927031040 |
|   | 3 | -0.899614482 |

Also, the plot of various energy states of Schrödinger for the WS potential versus diffuseness of the nuclear surface $a$ and the width of the potential $R$ are shown in Figures 3 and 4, respectively.

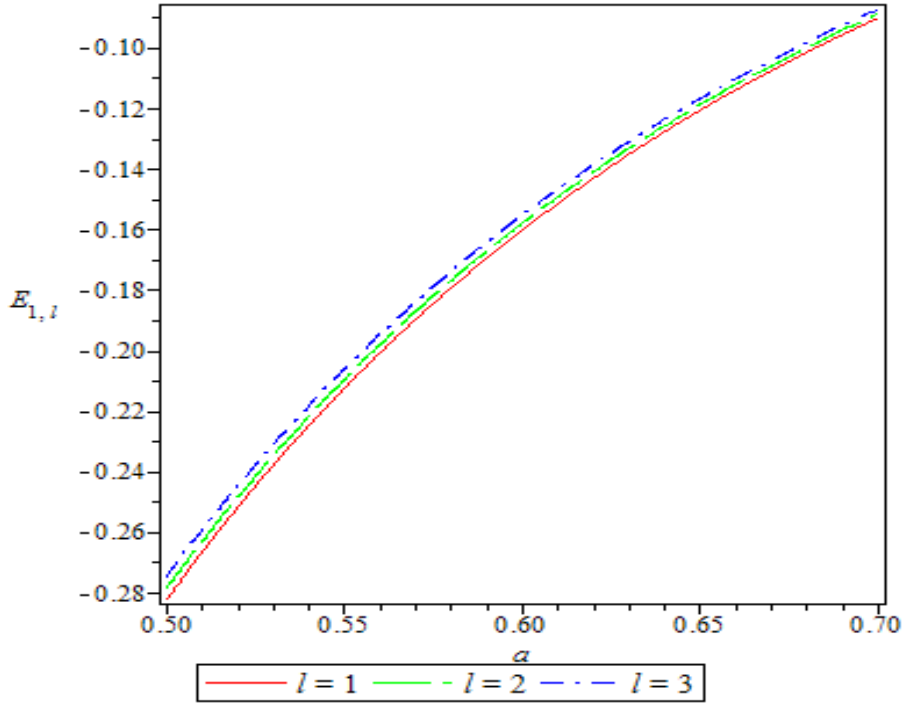

**Fig. 3:** Energy behavior of the Schrödinger equation with WS potential versus $a$ for various $l$.

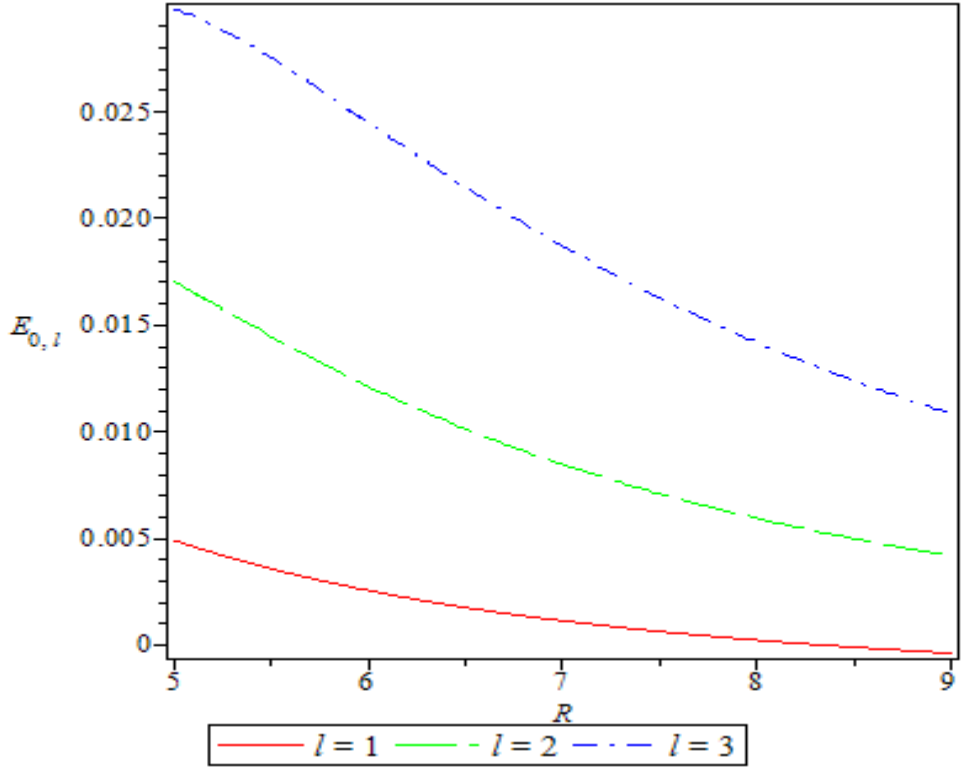

**Fig. 4:** Energy behavior of the Schrödinger equation with WS potential versus $R$ for various $l$.

Further, the non-relativistic wave function can be found as

$$\psi_{nl}(r) = B_{nl}\left(\frac{1}{1+e^{v(r-R)}}\right)^{\eta_l}\left(\frac{e^{v(r-R)}}{1+e^{v(r-R)}}\right)^{\lambda_i} P_n^{(2\eta_l, 2\lambda_i)}\left(\frac{e^{v(r-R)}-1}{e^{v(r-R)}+1}\right), \quad (24)$$

with

$$\lambda_i = \sqrt{\frac{l(l+1)}{v^2 R^2}\left(1+\frac{4}{vR}+\frac{12}{v^2 R^2}\right)-\frac{2\mu}{v^2\hbar^2}(E_{nl}+V_0)},$$

$$\eta_l = \sqrt{\frac{l(l+1)}{v^2 R^2}\left(1-\frac{4}{vR}+\frac{12}{v^2 R^2}\right)-\frac{2\mu E_{nl}}{v^2\hbar^2}}, \quad (25)$$

and $B_{nl}$ is the non-relativistic normalization factor. The exact $s$-wave energy and wave function solutions are obtained when $l = 0$ as

$$E_{nl} = \frac{\hbar^2 l(l+1)}{2\mu R^2}\left(1-\frac{4}{vR}+\frac{12}{v^2 R^2}\right)-\frac{\hbar^2}{2\mu(n-L)^2}\left(\frac{4l(l+1)}{v^2 R^3}-\frac{\mu V_0}{v\hbar^2}-\frac{1}{2}(n-L)^2\right)^2, \quad (26)$$

$$\psi_{nl}(r) = B_{nl}\left(\frac{1}{1+e^{v(r-R)}}\right)^{\eta_0}\left(\frac{e^{v(r-R)}}{1+e^{v(r-R)}}\right)^{\lambda_0} P_n^{(2\eta_0, 2\lambda_0)}\left(\frac{e^{v(r-R)}-1}{e^{v(r-R)}+1}\right), \quad (27)$$

where

$$\lambda_0 = \frac{1}{\nu}\sqrt{-\frac{2\mu}{\hbar^2}(E_{nl}+V_0)},\ E_{nl}<0,\qquad \eta_0 = \frac{1}{\nu}\sqrt{-\frac{2\mu E_{nl}}{\hbar^2}},\quad (28)$$

respectively.

## 4. Final remarks and conclusion

In this work, we have obtained approximate analytical solutions of the two-body spinless Salpeter equation with the usual WS potential interaction by using the parametric generalization of the NU method. The approximate semi-relativistic bound-state energy eigenvalues and corresponding wave functions are obtained by using Pekeris approximation to the centrifugal potential. The present solutions are valid when the diffuseness of the nuclear surface $a$ is large compared with $r-R$. Further, approximate nonrelativistic energy and wave function solutions for any $l$-state can be easily obtained under some limited conditions when $E_{nl} \ll \tilde{m}$.

## Acknowledgments

The authors wish to thank the kind referees for their invaluable suggestions which have greatly helped in the improvement of this paper.

## References

[1] Bethe H, Salpeter E. Phys. Rev., 1951, 84: 1232
[2] Nambu Y., Prog. Theor. Phys., 1950, 5: 614
[3] Wick G. C., Phys. Rev., 1954, 96: 1124
[4] Nakanishi N., Prog. Theor. Phys. Suppl., 1969, 43: 1
[5] Roberts C. D., Williams A. G., Prog. Part. Nucl. Phys., 1994, 33: 477
[6] Maris P., Roberts C. D., Phys. Rev. C, 1997, 56: 3369
[7] Roberts C. D., Schmidt S. M., Prog. Part. Nucl. Phys., 2000, 45: 1
[8] Maris P., Roberts C. D., Int. J. Mod. Phys. E, 2003, 12: 297
[9] Li Z. F., Lucha W., Schöberl F. F., J. Phys. G: Nucl. Part. Phys., 2008, 35: 115002
[10] Chang L., Roberts C. D., Phys. Rev. Lett., 2009, 103: 081601
[11] Lucha W., Mod. Phys. Lett. A, 1990, 5: 2473
[12] Lucha W., Schöberl F. F., Gromes D., Phys. Rep., 1991, 200: 127
[13] Lucha W., Schöberl F. F., Phys. Rev. A, 1999, 60: 5091

[14] Lucha W., Schöberl F. F., Phys. Rev. A, 1996, 54: 3790

[15] Lucha W., Schöberl F. F., Int. J. Mod. Phys. A, 1999, 14: 2309

[16] Lucha W., Schöberl F. F., Fizika B, 1999, 8: 193

[17] Lucha W., Schöberl F. F., Int. J. Mod. Phys. A, 2000, 15: 3221

[18] Hall R., Lucha W., Schöberl F. F., J. Phys. A: Math.Gen., 2001, 34: 5059

[19] Hall R., Lucha W., Schöberl F. F., Int. J. Mod. Phys. A, 2002, 17: 1931

[20] Lucha W., Schöberl F. F., Int. J. Mod. Phys. A, 2002, 17: 2233

[21] Hall R., Lucha W., Schöberl F. F., Int. J. Mod. Phys. A, 2003, 18: 2657

[22] Hall R., Lucha W., J. Phys. A: Math. Gen., 2005, 38: 7997

[23] Hall R., Lucha W., Int. J. Mod. Phys. A, 2007, 22: 1899

[24] Hall R., Lucha W., J. Phys. A: Math. Theor., 2008, 41: 355202

[25] Hall R., Lucha W., Phys. Lett. A, 2010, 374: 1980

[26] Ikhdair S. M., Sever R., Z. Phys. C, 1992, 56: 155

[27] Ikhdair S. M., Sever R., Z. Phys. C, 1993, 58: 153

[28] Jaczko G., Durand L., Phys. Rev. D, 1998, 58: 114017

[29] Ikhdair S. M., Sever R., Int. J. Mod. Phys. A, 2004, 19: 1771; Ikhdair S. M., Sever R., Int. J. Mod. Phys. A, 2005, 20: 6509; Ikhdair S. M., Sever R., Int. J. Mod. Phys. E, 2008, 17: 669; Ikhdair S. M., Sever R., Int. J. Mod. Phys. E, 2008, 17: 1107

[30] Tokunaga Y., 77th Annual Meeting of the South Eastern Section of the APS (20–23 October 2010) BAPS.2010.SES.CD.3

[31] Woods R. D., Saxon D. S., Phys. Rev., 1954, 95: 577

[32] Williams W. S. C. Nuclear and Particle Physics, Clarendon: Oxford, 1996.

[33] Garcia F., Garrote E., Yoneama M. L., et al. Eur. Phys. J. A, 1999, 6: 49

[34] Goldberg V. Z., Chubarian G. G., Tabacaru G., et al. Phys. Rev. C, 2004, 69: 031302(R)

[35] Syntfeld A., March H., Płóciennik W., et al. Eur. Phys. J. A, 2004, 20: 359

[36] Diaz-Torres A., Scheid W., Nucl. Phys. A, 2005, 757: 373

[37] Guo J. Y., Sheng Q., Phys. Lett. A, 2005, 338: 90

[38] Badalov V. H., Ahmadov H. I., Ahmadov A. I., Int. J. Mod. Phys. E, 2009, 18: 631

[39] Badalov V. H., Ahmadov H. I., Badalov S. V., Int. J. Mod. Phys. E, 2010, 18: 1463

[40] Chepurnov V. A., Nemirovsky P. E., Nucl. Phys., 1963, 49: 90

[41] Chasman R. R., Wilkins B. D., Phys. Lett. B, 1984, 149: 433

[42] Dudek J., Wemer T., J. Phys. G: Nuclear Phys., 1978, 4: 1543

[43] Nikiforov A. F., Uvarov V. B. Special Functions of Mathematical Physics, Birkhausr: Berlin, 1988.

[44] Ikhdair S. M., Int. J. Mod. Phys. C, 2009, 20: 1563

[45] Greene R. L., Aldrich C., Phys. Rev. A, 1976, 14: 2363

[46] Aydoğdu O., Sever R., Eur. Phys. J. A, 2010, 43: 73

**Appendix A: Parametric generalization of the NU method**

The NU method is used to solve the second-order differential equations with an appropriate coordinate transformation $s = s(r)$ [35]

$$\psi_n''(s) + \frac{\tilde{\tau}(s)}{\sigma(s)}\psi_n'(s) + \frac{\tilde{\sigma}(s)}{\sigma^2(s)}\psi_n(s) = 0, \tag{A1}$$

where $\sigma(s)$ and $\tilde{\sigma}(s)$ are polynomials, at most of the second degree, and $\tilde{\tau}(s)$ is a first-degree polynomial. To make the application of the NU method simpler and direct without the need to check the validity of solution. We present a short cut for the NU method. Therefore, at first we write the general form of the Schrödinger-like equation (A1) in a more general form applicable to any potential as follows [36,41-44]

$$\psi_n''(s) + \left(\frac{c_1 - c_2 s}{s(1 - c_3 s)}\right)\psi_n'(s) + \left(\frac{-As^2 + Bs - C}{s^2(1 - c_3 s)^2}\right)\psi_n(s) = 0, \tag{A2}$$

satisfying the wave functions

$$\psi_n(s) = \phi(s) y_n(s). \tag{A3}$$

Comparing (A2) with its counterpart (A1), we obtain the following identifications:

$$\tilde{\tau}(s) = c_1 - c_2 s, \quad \sigma(s) = s(1 - c_3 s), \quad \tilde{\sigma}(s) = -As^2 + Bs - C, \tag{A4}$$

Following the NU method [35], we obtain the followings [36,41-44],

(i) The relevant constant coefficients:

$$c_4 = \frac{1}{2}(1-c_1), \qquad c_5 = \frac{1}{2}(c_2 - 2c_3),$$

$$c_6 = c_5^2 + A, \qquad c_7 = 2c_4 c_5 - B,$$

$$c_8 = c_4^2 + C, \qquad c_9 = c_3(c_7 + c_3 c_8) + c_6,$$

$$c_{10} = c_1 + 2c_4 + 2\sqrt{c_8} - 1 > -1, \qquad c_{11} = 1 - c_1 - 2c_4 + \frac{2}{c_3}\sqrt{c_9} > -1,\ c_3 \neq 0,$$

$$c_{12} = c_4 + \sqrt{c_8} > 0, \qquad c_{13} = -c_4 + \frac{1}{c_3}(\sqrt{c_9} - c_5) > 0,\ c_3 \neq 0. \tag{A5}$$

(ii) The essential polynomial functions:

$$\pi(s) = c_4 + c_5 s - \left[\left(\sqrt{c_9} + c_3 \sqrt{c_8}\right)s - \sqrt{c_8}\right], \tag{A6}$$

$$k = -(c_7 + 2c_3 c_8) - 2\sqrt{c_8 c_9}, \tag{A7}$$

$$\tau(s) = c_1 + 2c_4 - (c_2 - 2c_5)s - 2\left[\left(\sqrt{c_9} + c_3 \sqrt{c_8}\right)s - \sqrt{c_8}\right], \tag{A8}$$

$$\tau'(s) = -2c_3 - 2\left(\sqrt{c_9} + c_3 \sqrt{c_8}\right) < 0. \tag{A9}$$

(iii) The energy eigenvalue calculations

$$(c_2 - c_3)n + c_3 n^2 - (2n+1)c_5 + (2n+1)\left(\sqrt{c_9} + c_3 \sqrt{c_8}\right) + c_7 + 2c_3 c_8 + 2\sqrt{c_8 c_9} = 0. \tag{A10}$$

(iv) The wave function calculations

$$\rho(s) = s^{c_{10}}(1 - c_3 s)^{c_{11}}, \tag{A11}$$

$$\phi(s) = s^{c_{12}}(1 - c_3 s)^{c_{13}},\ c_{12} > 0,\ c_{13} > 0, \tag{A12}$$

$$y_n(s) = P_n^{(c_{10}, c_{11})}(1 - 2c_3 s),\ c_{10} > -1,\ c_{11} > -1, \tag{A13}$$

$$\psi_{n\kappa}(s) = N_{n\kappa} s^{c_{12}}(1 - c_3 s)^{c_{13}} P_n^{(c_{10}, c_{11})}(1 - 2c_3 s). \tag{A14}$$

where $P_n^{(\mu,\nu)}(x)$, $\mu > -1$, $\nu > -1$, and $x \in [-1,1]$ are Jacobi polynomials with

$$P_n^{(\alpha,\beta)}(1 - 2s) = \frac{(\alpha+1)_n}{n!}\ {}_2F_1(-n, 1 + \alpha + \beta + n; \alpha + 1; s), \tag{A15}$$

and $N_{n\kappa}$ is a normalization constant. Also, the above wave functions can be expressed in terms of the hypergeometric function as

$$\psi_{n\kappa}(s) = N_{n\kappa} s^{c_{12}}(1 - c_3 s)^{c_{13}}\ {}_2F_1(-n, 1 + c_{10} + c_{11} + n; c_{10} + 1; c_3 s) \tag{A16}$$

where $c_{12} > 0$, $c_{13} > 0$ and $s \in [0, 1/c_3]$, $c_3 \neq 0$.